\documentclass[a4paper]{jpconf}
\usepackage{graphicx}
\usepackage{amsmath}
\usepackage{multirow}
\usepackage{cite}

\begin{document}
\title{$X(3872)$ production from reactions involving $D$ and $D^*$ mesons}
\author{A.~Mart\'inez~Torres$^1$, K. P. Khemchandani$^1$, F. S. Navarra$^1$, M. Nielsen$^1$, Luciano M. Abreu$^2$}
\address{$^1$ Instituto de F\'isica, Universidade de S\~ao Paulo, C.P. 66318, 05389-970 S\~ao 
Paulo, SP, Brazil.}
\address{$^2$ Instituto de F\'isica, Universidade Federal da Bahia, 40210-340, Salvador, BA, Brazil.}

\begin{abstract}
In this proceeding we show the results found for the cross sections of the processes $\bar D D\to\pi X(3872)$, $\bar D^* D\to \pi X(3872)$ and $\bar D^* D^*\to\pi X(3872)$, information needed for calculations of the $X(3872)$ abundance in heavy ion collisions. Our formalism is based on the generation of $X(3872)$ from the interaction of the hadrons $\bar D^0 D^{*0} - \textrm{c.c}$, $D^- D^{*+} - \textrm{c.c}$ and $D^-_s D^{*+}_s - \textrm{c.c}$. The evaluation  of the cross section associated with processes  having $D^*$ meson(s) involves an anomalous vertex, $X\bar D^* D^*$, which we have determined by considering triangular loops motivated by the molecular nature of $X(3872)$.  We find that the contribution of this vertex is important. Encouraged by this finding we estimate the $X\bar D^* D^*$ coupling, which turns out to be $1.95\pm 0.22$. We then use it to obtain the cross section for the reaction $\bar D^* D^*\to\pi X$ and find that the $X\bar D^* D^*$ vertex is also relevant in this case. We also discuss the role of the charged components of $X$ in the determination of the production cross sections.
\end{abstract}

\section{Introduction}
Since the development of $B$ factories like BELLE and BES a wealth of data on new hadronic states has been produced~\cite{Uchida,Guo}, information which is crucial to understand the nature and properties of such states. Particularly interesting are the data on the so called exotic charmonium states. One member of this family, and probably the most widely studied theoretically, is the $X(3872)$ (from now on simply $X$),  reported a decade ago by the Belle collaboration in the decay $B^{\pm}\to K^{\pm}\pi^+\pi^- J/\psi$~\cite{Choi}. After this finding,  several other collaborations  
\cite{Acosta,Abazov,Aubert} confirmed this state and its existence is now established beyond any doubt.  However, it has only been very recently when the spin-parity quantum numbers of $X$ have been confirmed to be  
$1^{++}$~\cite{Aaij}. 

During these years, several theoretical models have been proposed  to describe the properties of this state, considering it as a charmonium state, a tetraquark, a $D - \bar{D^*}$ hadron molecule  
and a mixture between a charmonium and a molecular component~\cite{Tornqvist,Close,Swanson,Braaten,Daniel2,Nielsen,Matheus, Dong,Daniel3,Daniel4,Dubnicka,Badalian,Coito,roma14}. In spite of the effort of these numerous groups, the properties of this particle are not yet well understood and represent a challenge both for theorists and experimentalists.  

In line with the information on new charmonium states brought by the $B$ factories, in a different frontier of physics, collaborations like RHIC and LHC has devoted a significant part of their physics program to study the Quark Gluon Plasma (QGP). It is now a well accepted fact that in high energy heavy ion collisions a deconfined medium is created:  the quark gluon plasma (QGP) \cite{Arsene,Adams}. In a high energy heavy ion collision the QGP is formed, expands, cools, hadronizes and is converted into a hadron gas, which lives up to $10$ fm/c and then freezes out. During this evolution, an increasing (with the reaction energy) number of charm quarks and anti-quarks move freely. The initially formed charmonium bound states are dissolved (the famous ``charmonium suppression") but $c$'s and $\bar{c}$'s, coming now from different parent gluons, can pick up light quarks and anti-quarks from the rich environment and form multiquark bound states. This is called quark coalescence and it happens during the phase transition to the hadronic gas \cite{Cho1,Cho2}.  Therefore, the formation of the quark gluon plasma phase increases the number of produced $X$'s \cite{Cho1,Cho2}. Interestingly, the coalescence formalism is based on the overlap of the Wigner functions of the quarks and of the bound state, being thus sensitive to the spatial configuration of the charmonium state and hence being able to distinguish between a compact, $\simeq 1$ fm long, tetraquark configuration and a large $\simeq 10$ fm long, molecular configuration. A big difference between the predicted abundancies could be used as a tool to discriminate between different  $X$ structures and to help us to decide whether it is a molecule or a tetraquark \cite{Cho3}.  
In this way, heavy ion collisions can be used to obtain information about exotic charmonium states as $X$. However, there is an additional complication. Due to the rich hadronic environment  present in the plasma,
the $X$'s can be destroyed in collisions with ordinary hadrons, such as $X + \pi \rightarrow D + \bar{D^*}$, and can also be produced through the inverse reactions, such as $D + \bar{D}^* \rightarrow X + \pi$.  A proper determination of the abundance of $X$ in heavy ion collisions requires a precise calculation of the cross sections of these kind of processes. In Ref.~\cite{Cho3}, the hadronic absorption cross section of the $X$ by mesons like $\pi$ and $\rho$ was evaluated  for the processes $\pi X \to D\bar D$, $\pi X \to D^*\bar D^*$,  $\rho X\to D\bar D$,  $\rho X \to D\bar D^*$, and $\rho X \to D^*\bar D^*$. Using these  cross sections, the variation of the $X$ meson abundance during the expansion of the hadronic matter was computed with the help of a kinetic equation with gain and loss terms. The results turned out to be  strongly dependent on the quantum numbers of the $X$ and on its structure.

The present work is devoted to introduce two improvements in the calculation of cross sections performed in Ref.~\cite{Cho3}.  
The first and most important one is the inclusion of the anomalous vertices $\pi D^* D^*$ and $X \bar D^* D^*$, which were neglected 
before. With these vertices new reaction channels become possible, such as  $\pi X\to D\bar D^*$, and the inverse process
$D\bar D^* \to \pi X$. As will be seen, this reaction is the most important one for $X$ in the hadron gas. 
The relevance of anomalous couplings has also been shown earlier in different contexts, for example in the $J/\psi$ 
absorption cross sections by $\pi$ and $\rho$ mesons~\cite{Oh}, radiative decays of scalar resonances and axial vector 
mesons~\cite{Nagahiro1,Nagahiro2} and in kaon photoproduction~\cite{Ozaki}.

The second  improvement is the inclusions of the charged components of the $D$ and $D^*$ mesons which couple to the $X$~\cite{Daniel4}. 

\section{Formalism}
\subsection{Determination of the cross sections}
To calculate the cross section for the processes (1) $\bar D D \to \pi X$,  (2) $\bar D^* D \to \pi X$ and (3) $\bar D^* D^*\to \pi X$ we consider the model of Refs.~\cite{Daniel4,Daniel3,Francesca2} in which $X$ is generated from the interaction of $\bar D^0 D^{*0}-\textrm{c.c}$, $D^- D^{*+}-\textrm{c.c}$ and $D^-_s D^{*+}_s-\textrm{c.c}$.  The isospin-spin averaged production cross section for the processes $\bar D D, \bar D^* D, \bar D^* D^*\to \pi X$, in the center of mas (CM) frame can be determined as
\begin{align}
\sigma_r(s)=\frac{1}{16\pi\lambda(s,m^2_{1i,r},m^2_{2i,r})}\int^{t_{\textrm{max,r}}}_{t_\textrm{min,r}}dt\overline{\sum\limits_{\textrm{Isos},\textrm{spin}}}\left |\mathcal{M}_r(s,t)\right|^2,\label{cross}
\end{align}
where $r=1,2,3$ is an index indicating the reaction considered, $\sqrt{s}$ is the CM energy, and $m_{1i,r}$ and $m_{2i,r}$ represent the masses of the two
particles present in the initial state $i$ of the reaction $r$. We follow the convention of associating the index 1 (2) with the particle with charm $-1$ ($+1$) present in the initial state. The function $\lambda(a,b,c)$ in Eq.~(\ref{cross}) is the K\"allen function, $t_\textrm{min,r}$ and $t_{\textrm{max,r}}$ correspond to the minimum and maximum values, respectively, of the Mandelstam variable $t$ and $\mathcal{M}_r$ is the reduced matrix element for the process $r$.  The symbol $\overline{\sum\limits_{\textrm{spin},\textrm{Isos}}}$ represents the sum over the isospins and spins of the particles in the initial and final state, weighted by the isospin and spin degeneracy factors of the two particles forming the initial state for the reaction $r$, i.e.,
\begin{align}
\overline{\sum\limits_{\textrm{spin},\textrm{Isos}}}\left|\mathcal{M}_r\right|^2\to \frac{1}{(2I_{1i,r}+1)(2I_{2i,r}+1)}\frac{1}{(2s_{1i,r}+1)(2s_{2i,r}+1)}\sum\limits_{\textrm{spin},\textrm{Isos}}\left|\mathcal{M}_r\right|^2,
\end{align}
where,
\begin{align}
\sum\limits_{\textrm{spin},\textrm{Isos}}\left|\mathcal{M}_r\right|^2=\sum\limits_{Q_{1i},Q_{2i}}\left[\sum\limits_{\textrm{spin}}\left|\mathcal{M}^{(Q_{1i},Q_{2i})}_r\right|^2\right].\label{Mqq}
\end{align}
In Eq.~(\ref{Mqq}), $Q_{1i}$ and $Q_{2i}$ represent the charges for each of the two particles forming the initial state $i$ of the reaction $r$, which are combined to obtain total charge $Q_r=Q_{1i}+Q_{2i}=0,+1,-1$. In this way,
we have four possibilities: $(0,0)$, $(-,+)$, $(-,0)$ and $(0,+)$ and thus,
\begin{align}
\sum\limits_{\textrm{spin},\textrm{Isos}}\left|\mathcal{M}_r\right|^2=\sum\limits_{\textrm{spin}}\left(\left|\mathcal{M}^{(0,0)}_r\right|^2+\left|\mathcal{M}^{(-,+)}_r\right|^2+\left|\mathcal{M}^{(-,0)}_r\right|^2+\left|\mathcal{M}^{(0,+)}_r\right|^2\right).
\end{align}.

Each of the amplitudes $\mathcal{M}^{(Q_{1i},Q_{2i})}_r$ of Eq.~(\ref{Mqq}) can be written as
\begin{align}
\mathcal{M}^{(Q_{1i},Q_{2i})}_r=T^{(Q_{1i},Q_{2i})}_r+U^{(Q_{1i},Q_{2i})}_r,
\end{align}
where $T^{(Q_{1i},Q_{2i})}_r$ and $U^{(Q_{1i},Q_{2i})}_r$ are the contributions related to the $t$ and $u$ channel diagrams contributing to each process.

\subsection{The $\bar D D \to \pi X$ reaction}
\begin{figure}[h!]
\centering
\includegraphics[width=0.7\textwidth]{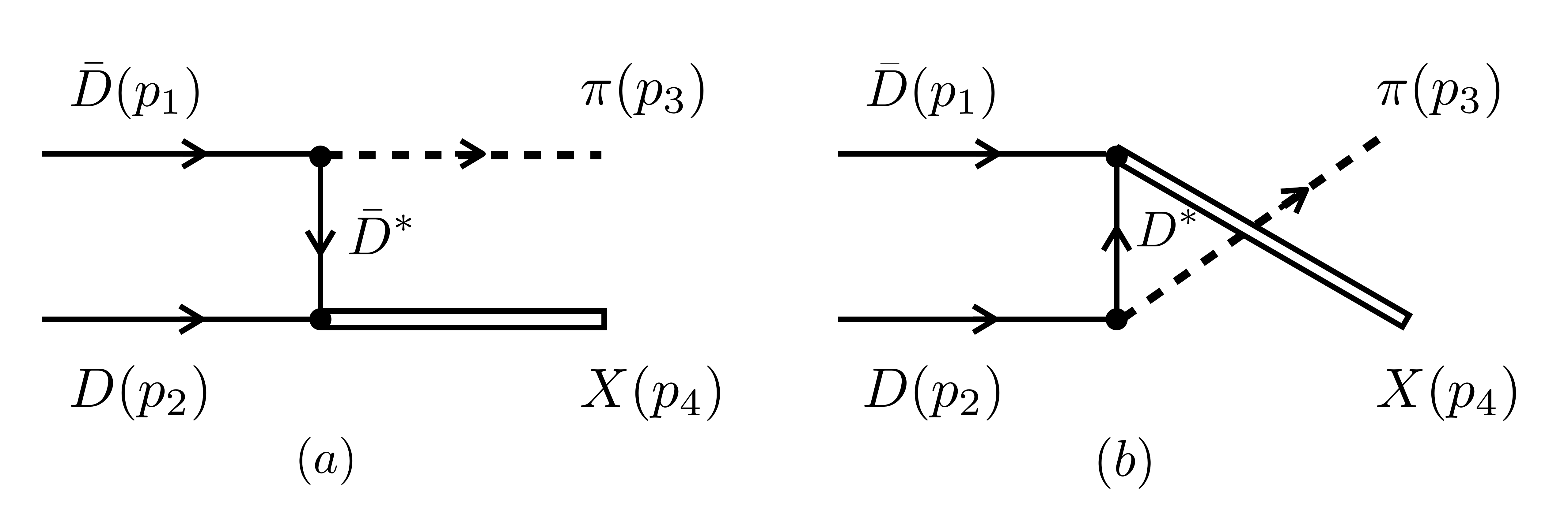}
\caption{Diagrams contributing to the process $\bar D D\to \pi X$.}\label{DbarD}
\end{figure}

In Fig.~\ref{DbarD} we show the different diagrams contributing to $\bar D D\to \pi X$ (without specifying the charge of the reaction).

The $t$-channel amplitude for the process in Fig.~\ref{DbarD}a can be written as
\begin{align}
T^{(Q_{1i},Q_{2i})}_1=W^{(Q_{1i},Q_{2i})}_1\,g_{PPV}\,g_{X}\frac{1}{t-m^2_{\bar{D}^*}}\left[(p_1+p_3)_\mu+\frac{m^2_{\bar{D}}-m^2_\pi}{m^2_{\bar{D}^*}}p_{2\mu}\right]\epsilon^\mu_X(p_4),\label{Tampl1}
\end{align}
while for the $u$-channel amplitude (Fig.~\ref{DbarD}b) we have
\begin{align}
U^{(Q_{1i},Q_{2i})}_1=Z^{(Q_{1i},Q_{2i})}g_{PPV} g_X\frac{1}{u-m^2_{D^*}}\left[(p_2+p_3)_\mu+\frac{m^2_D-m^2_\pi}{m^2_{D^*}}p_{1\mu}\right]\epsilon^\mu_X(p_4).\label{Uampl}
\end{align}
The coefficients $W^{(Q_{1i},Q_{2i})}_r$ and $Z^{(Q_{1i},Q_{2i})}$ and couplings $g_X$ are given in Tables~\ref{Ccoeff} and \ref{Zcoeff}.

\begin{table}[h!]
\centering
\caption{Coefficients $W^{(Q_{1i},Q_{2i})}_r$ and couplings $g_X$ for the amplitude given in Eq.~(\ref{Tampl1}). We have defined $g_n\equiv g_{X\bar D^0 D^{*0}}$ and $g_c\equiv g_{X D^- D^{*+}}$,
whose numerical values can be found in Table~\ref{tableX}.\\}\label{Ccoeff}
\begin{tabular}{cccc}
\hline\hline
$r$\quad\quad&$(Q_{1i},Q_{2i})$&$W_r$\quad\quad&$g_X$\\
\hline
\multirow{2}{*}{1}\quad\quad&$(0,0)$&$-1/\sqrt{2}$\quad\quad&$-g_n$\\&$(-,+)$&$1/\sqrt{2}$\quad\quad &$-g_c$\\
&$(-,0)$&$-1$\quad\quad&$-g_n$\\
&$(0,+)$&$-1$\quad\quad&$-g_c$\\
\hline
\multirow{2}{*}{2}\quad\quad&$(0,0)$&$-1/2$\quad\quad&$-g_n$\\&$(-,+)$&$1/2$\quad\quad &$-g_c$\\
&$(-,0)$&$-1/\sqrt{2}$\quad\quad&$-g_n$\\
&$(0,+)$&$-1/\sqrt{2}$\quad\quad&$-g_c$
\end{tabular}
\end{table}

\begin{table}[h!]
\centering
\caption{Coefficients $Z^{(Q_{1i},Q_{2i})}$ and couplings $g_X$ for the amplitude given in Eq.~(\ref{Uampl}). We have defined $g_n\equiv g_{X\bar D^0 D^{*0}}$ and $g_c\equiv g_{XD^- D^{*+}}$,
whose numerical values can be found in Table~\ref{tableX}.\\}\label{Zcoeff}
\begin{tabular}{ccc}
\hline\hline
$(Q_{1i},Q_{2i})$&$Z_r$\quad\quad&$g_X$\\
\hline
$(0,0)$&$1/\sqrt{2}$\quad\quad&$g_n$\\
$(-,+)$&$-1/\sqrt{2}$\quad\quad &$g_c$\\
$(-,0)$&$1$\quad\quad&$g_c$\\
$(0,+)$&$1$\quad\quad&$g_n$\\
\end{tabular}
\end{table}

\begin{table}[h!]
\centering
\caption{Couplings of $X$ to the different pseudoscalar-vector components constituting the state ($\bar{P}_X V_X$). The couplings for the complex conjugate components bear a minus sign.\\}
\label{tableX}
\begin{tabular}{c c}
\hline\hline
$\bar{P}_X V_X$&$g_{X\bar{P}_X V_X}$ (MeV)\\
\hline
$D^- D^{*+}$ & $3638/\sqrt{2}$ \\
$\bar{D}^0 D^{*0}$&$3663/\sqrt{2}$ \\
$D^-_s D^{*+}_s$ &$3395/\sqrt{2}$
\end{tabular}
\end{table}

The coupling $g_{PPV}$  in Eqs.~(\ref{Tampl1}) and (\ref{Uampl}) is the strong coupling of the $D^*$ meson to $D\pi$. As shown in Refs.~\cite{Liang,Francesca}, consideration of heavy quark symmetry gives a value of
\begin{align}
g_{PPV}=\frac{m_\rho}{2 f_\pi}\frac{m_{D^*}}{m_{K^*}}~\sim 9,\label{HQS}
\end{align}
where we have use the pion decay constant value $f_\pi=93$ MeV. Using this coupling, the decay width for the process $D^{*+}\to D^0\pi^+$ is 71 KeV, in agreement with the recent experimental result of $(65\pm 15)$ KeV~\cite{Anastassov} and compatible with the coupling found in Ref.~\cite{Bracco} using QCD sum rules.

\subsection{The $\bar D^* D\to\pi X$ reaction}
\begin{figure}[h!]
\centering
\includegraphics[width=0.65\textwidth]{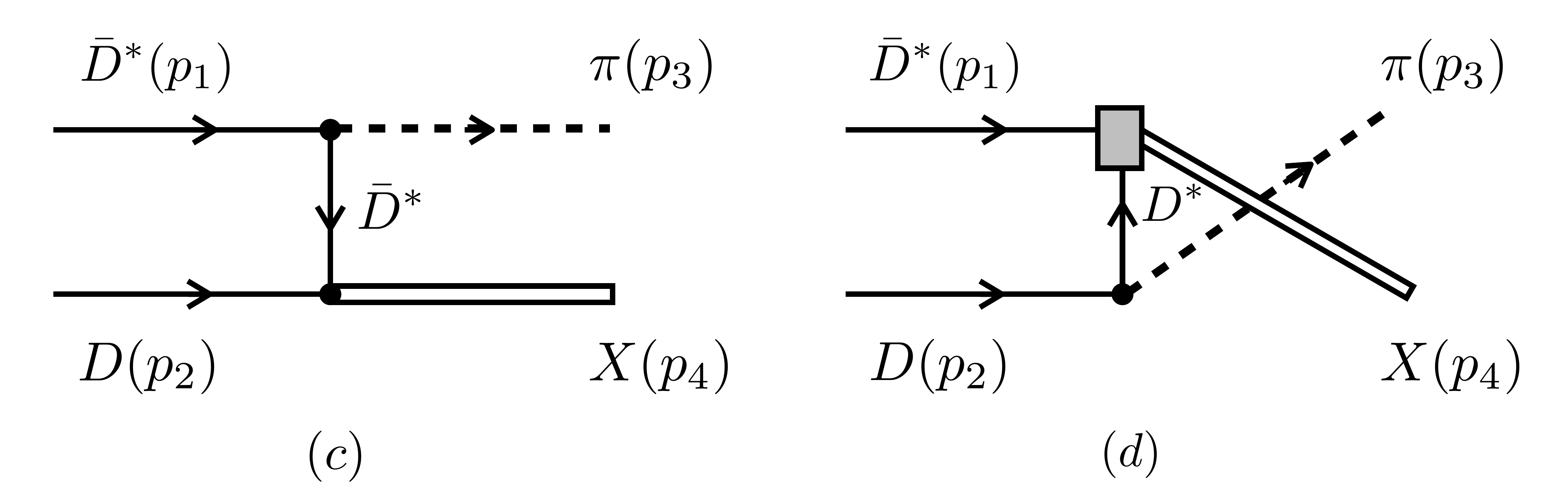}
\caption{Diagrams contributing to the process $\bar D^* D\to \pi X$. The diagram containing a filled box is calculated by summing the set of diagrams shown in Fig.~\ref{blob}, as explained in the text.}\label{DbarstarD}
\end{figure}

\begin{figure}[h!]
\centering
\includegraphics[width=0.73\textwidth]{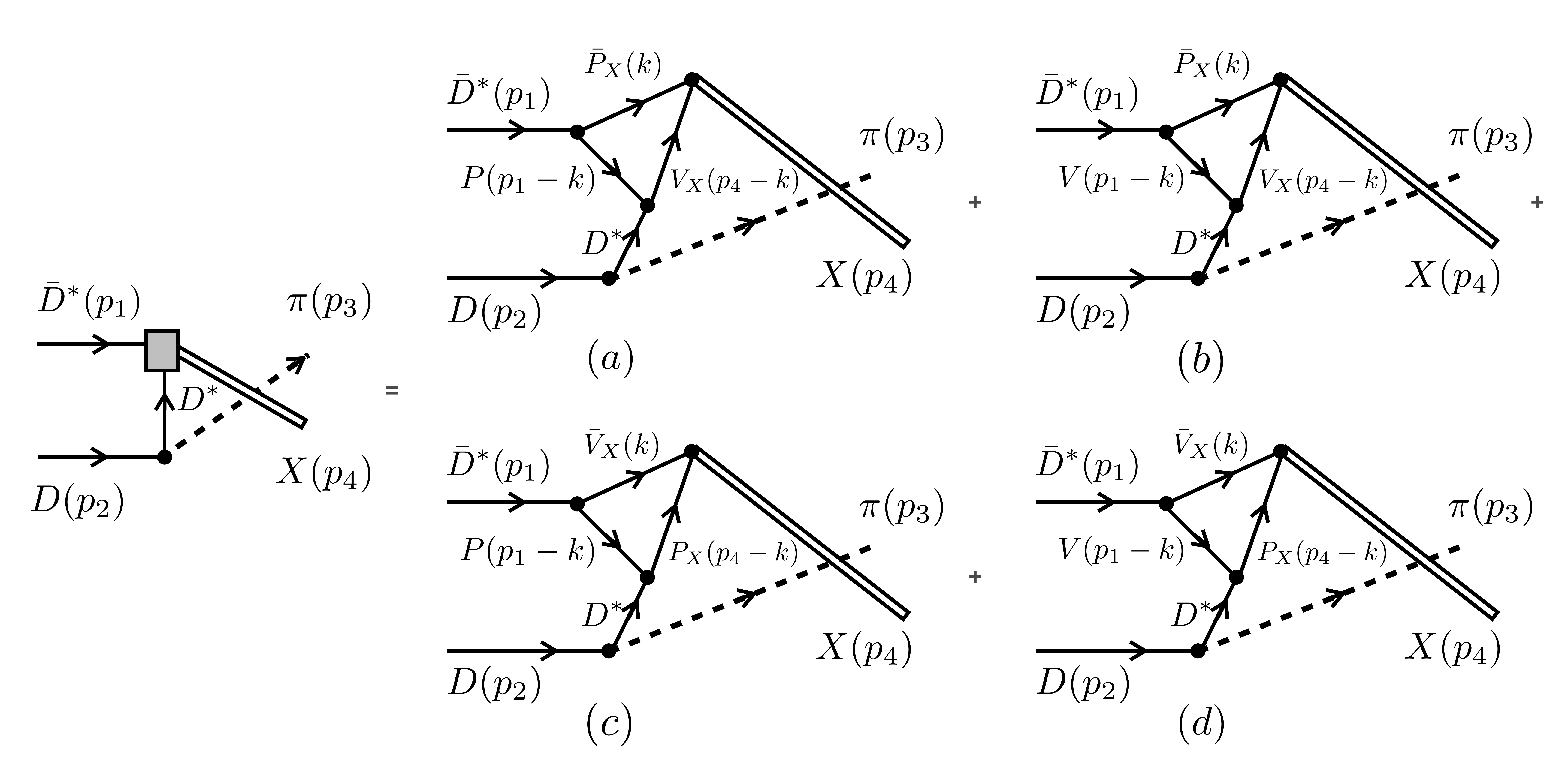}
\caption{Diagrams considered for the determination of the filled box shown in Fig.~\ref{DbarstarD}. The hadrons $P_X$ and $V_X$ represent the pseudoscalars and vectors coupling to the state $X$, while $P$ and $V$ are any
pseudoscalar and vector meson which can be exchanged conserving different quantum numbers. For a list of the different exchanged hadrons considered here see Ref.~\cite{our}.}\label{blob}
\end{figure}

We show the relevant diagrams contributing to this process in Figs.~\ref{DbarstarD} and \ref{blob}, which involve anomalous vertices, $\bar D^* \bar D^*\pi$ in the $t$-channel and $X\bar D^* D^*$ in the $u$-channel. 
The $t$ channel contribution is directly obtained as
\begin{align}
T^{(Q_{1i},Q_{2i})}_2=W^{(Q_{1i},Q_{2i})}_2\,g_{VVP}\,g_{X}\frac{1}{t-m^2_{\bar{D}^*}}\epsilon^{\mu\nu\alpha\beta}p_{1\mu}p_{3\alpha}\epsilon_{\bar D^*\nu}(p_1)\epsilon_{X\beta}(p_4).\label{Tampl2}
\end{align}
where the coefficients $W^{(Q_{1i},Q_{2i})}$ and the corresponding $g_X$ coupling are listed in Table~\ref{Ccoeff}. The amplitudes for the $u$-channel diagram shown in Fig.~\ref{DbarstarD}d can be calculated as
\begin{align}
U^{(Q_{1i},Q_{2i})}_2=\sum\limits_{p=a}^d U^{(Q_{1i},Q_{2i})}_{2p},
\end{align}
with $U^{(Q_{1i},Q_{2i})}_{2p}$ ($p=a,b,\cdots,d$) being the amplitudes associated with the diagrams depicted in Fig.~\ref{blob}. As can be seen, these amplitudes depend on the hadrons present in the triangular loops ($P$, $V_X$, etc.), since the couplings, propagators, etc., depend on them. The final result for the amplitude of each diagram in Fig.~\ref{blob} can be obtained by summing over the amplitudes for the different intermediate states 
\begin{align}
U^{(Q_{1i},Q_{2i})}_{2p}=\sum\limits_{P,P_X, V_X,V}\mathcal{U}^{(Q_{1i},Q_{2i})}_{2p},\label{Usum}
\end{align}
where $\mathcal{U}^{(Q_{1i},Q_{2i})}_{2p}$, $p=a,b,\textrm{etc.}$, is the amplitude for the diagram in Fig.~\ref{blob}p for a particular set of hadrons in the triangular loop.

The evaluation of the amplitudes in Eqs.~(\ref{Tampl2}) and (\ref{Usum}) involves $PPV$, $VVP$ and $VVV$ vertices (with $P$ and $V$ representing a pseudoscalar and a vector meson, respectively). To calculate them we have made used of effective Lagrangians~\cite{Bando1,Bando2,Meissner,Harada}
\begin{align}
\mathcal{L}_{PPV}&=-i g_{PPV}  \langle V^\mu [P,\partial_\mu P]\rangle,\nonumber\\
\mathcal{L}_{VVP}&=\frac{g_{VVP}}{\sqrt{2}}\epsilon^{\mu\nu\alpha\beta}\langle\partial_\mu V_\nu\partial_\alpha V_\beta P\rangle\label{Lag}\\
\mathcal{L}_{VVV}&=i g_{VVV} \langle(V^\mu\partial_\nu V_\mu-\partial_\nu V_\mu V^\mu) V^\nu)\rangle.\nonumber
\end{align}
with 
\begin{align}
g_{VVP}=\frac{3 m^2_V}{16 \pi^2 f^3_\pi},\quad
g_{VVV}=\frac{m_V}{2 f_\pi},\label{coup}
\end{align}
The symbol $\langle\,\rangle$  in Eq.~(\ref{Lag}) indicates the trace in the isospin space. 

The determination of the amplitudes in Eq.~(\ref{Usum}) is quite tedious and we refer to the reader to Ref.~\cite{our} for more details on the calculations and for a list of the different intermediate channels considered.

A different way to proceed in the determination of the $u$-channel diagram in Fig.~\ref{DbarstarD}d is to construct an effective Lagrangian of the type~\cite{Maiani}
\begin{align}
\mathcal{L}_{X \bar D^* D^*}=i g_{X\bar D^* D^*} \epsilon^{\mu\nu\alpha\beta}\partial_\mu X_\nu \bar D^*_\alpha D^*_\beta,\label{effL}
\end{align}
and try to estimate somehow the unknown coupling $g_{X\bar D^* D^*}$. However, a model like this would lose its predictive power in the absence of any reasonable constrain on the value of the coupling  $g_{X\bar D^* D^*}$. The strategy followed in this paper consists of first determining the $\bar D^* D\to\pi X$ cross section by calculating the $X\bar D^* D^*$ vertex in terms of the loops shown in Fig.~\ref{blob}. After this is done, we obtain the cross section for the same process but using the Lagrangian in Eq.~(\ref{effL}) to evaluate the diagram in Fig.~\ref{DbarstarD}d and compare both results. In this way, we get a reliable estimation of the  $g_{X\bar D^* D^*}$ coupling.

\subsection{The $\bar D^* D^*\to\pi X$ reaction}
\begin{figure}[h!]
\centering
\includegraphics[width=0.65\textwidth]{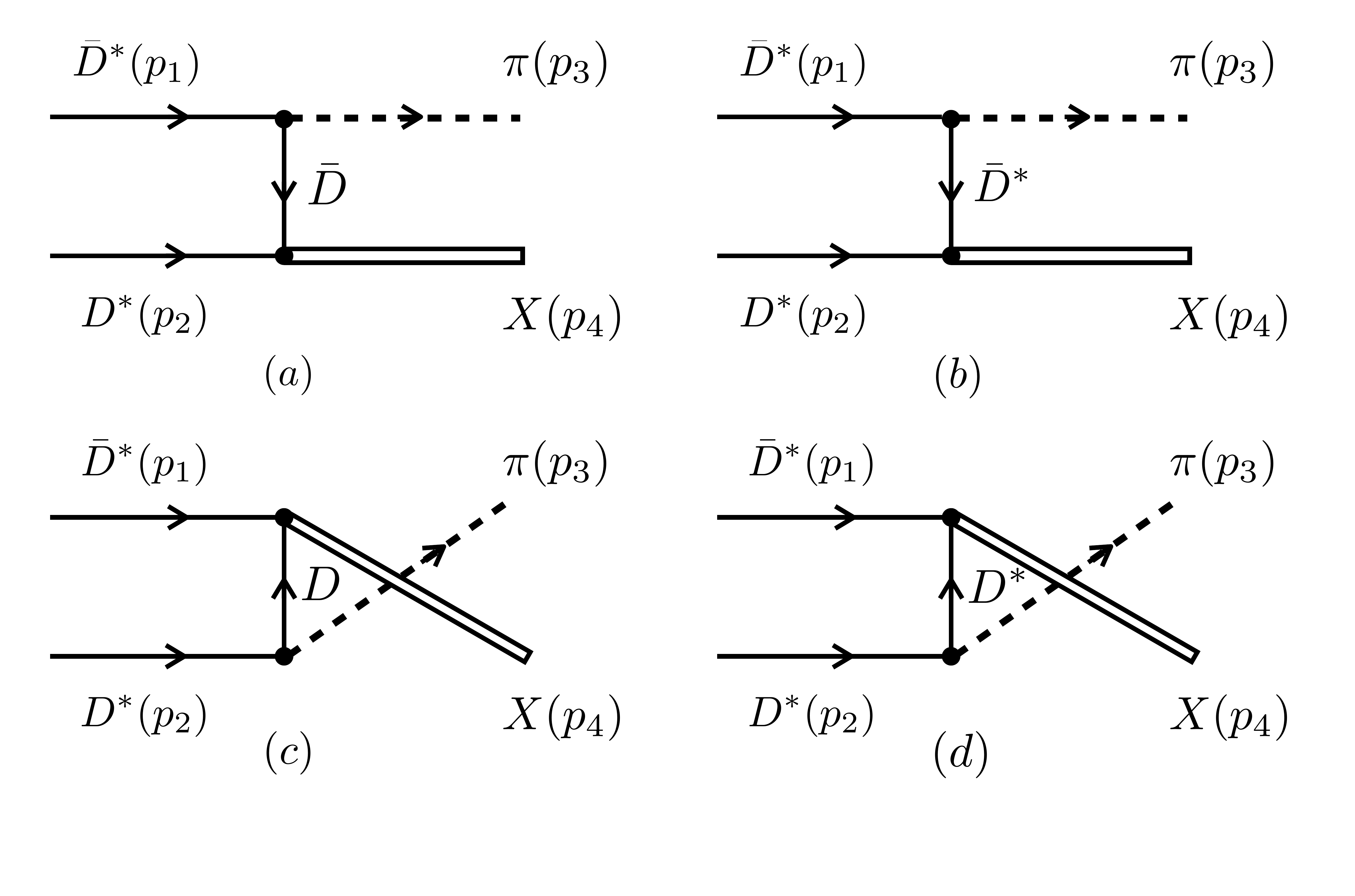}
\caption{Different diagrams contributing to the reaction $\bar D^* D^*\to\pi X$.}\label{DstarDstar}
\end{figure}

As shown in Fig.~\ref{DstarDstar}, the cross section for the process $\bar D^* D^*\to\pi X$  can get contributions from the anomalous $X\bar D^* D^*$ vertex. To determine the diagrams in Figs.~\ref{DstarDstar}b and \ref{DstarDstar}d we are going to make use of the method explained in the previous section to estimate the coupling $g_{X\bar D^* D^*}$ and consider the $X\bar D^* D^*$ vertex as a point-like one. 
Considering the Lagrangian in Eq.~(\ref{effL}) for the $X\bar D^* D^*$ vertex, we find the following amplitudes for the $t$ and $u$ channel diagrams:
\begin{align}
T^{(Q_{1i},Q_{2i})}_{3a}&=-2 g_{PPV}\,g^a_{X}\,\mathcal{Y}^{(Q_{1i},Q_{2i})}\frac{1}{t-m^2_{\bar D}}p_{3\mu}\epsilon^\mu_{\bar D^*}(p_1)\epsilon^\nu_{D^*}(p_2)\epsilon_{X\nu}(p_4)\nonumber\\
T^{(Q_{1i},Q_{2i})}_{3b}&=-\frac{g_{VVP}}{\sqrt{2}}g_{X\bar D^* D^*}\,\mathcal{Y}^{(Q_{1i},Q_{2i})}\frac{1}{t-m^2_{\bar D^*}}\,\epsilon^{\mu\nu\alpha\beta}\,\epsilon^{\mu^\prime\nu^\prime\alpha^\prime}_{\phantom{\alpha}\phantom{\mu}\phantom{\nu}\phantom{\beta}\beta} \,p_{1\mu}\,p_{3\alpha}\,p_{4\mu^\prime}\epsilon_{\bar D^*\nu}(p_1)\epsilon_{D^*\alpha^\prime}(p_2)\epsilon_{X\nu^\prime}(p_4)\label{TUDstar}\\
U^{(Q_{1i},Q_{2i})}_{3c}&=-2 g_{PPV}\,g^c_{X}\,\mathcal{Y}^{(Q_{1i},Q_{2i})}\frac{1}{u-m^2_{D}}\,p_{3\nu}\,\epsilon^\mu_{\bar D^*}(p_1)\epsilon^\nu_{D^*}(p_2)\epsilon_{X\mu}(p_4)\nonumber\\
U^{(Q_{1i},Q_{2i})}_{3d}&=- \frac{g_{VVP}}{\sqrt{2}}g_{X\bar D^* D^*}\,\mathcal{Y}^{(Q_{1i},Q_{2i})}\frac{1}{u-m^2_{D^*}}\epsilon^{\mu\nu\alpha\beta}\epsilon^{\mu^\prime\nu^\prime\alpha^\prime\beta^\prime} g_{\nu^\prime\alpha}p_{2\alpha^\prime}p_{3\mu^\prime}p_{4\mu}\epsilon_{\bar D^*\beta}(p_1)\epsilon_{D^*\beta^\prime}(p_2)\epsilon_{X\nu}(p_4),\nonumber
\end{align}
where the values of $g^a_X$, $g^c_X$, and $\mathcal{Y}^{(Q_{1i},Q_{2i})}$ are those given in Table~\ref{Ycoeff}.
\begin{table}
\centering
\caption{Values for the coupling $g^{a,c}_X$ and the coefficients $\mathcal{Y}^{(Q_{1i},Q_{2i})}$ of Eq.~(\ref{TUDstar}). The numerical values of $g_n$ and $g_c$ can be found in Table~\ref{tableX}.\\}\label{Ycoeff}
\begin{tabular}{cccc}
\hline\hline
$(Q_{1i},Q_{2i})$&$g^a_X$&$g^c_X$&$\mathcal{Y}^{(Q_{1i},Q_{2i})}$\\
\hline
$(0,0)$&$g_n$&$g_n$&$\frac{1}{\sqrt{2}}$\\
$(-,+)$&$g_c$&$g_c$&$-\frac{1}{\sqrt{2}}$\\
$(-,0)$&$g_n$&$g_c$&$1$\\
$(0,+)$&$g_c$&$g_n$&$1$
\end{tabular}
\end{table}
\section{Results}\label{Res}
\subsection{The $\bar D D\to \pi X$ reaction}
\begin{figure}[t]
\centering
\includegraphics[width=0.4\textwidth]{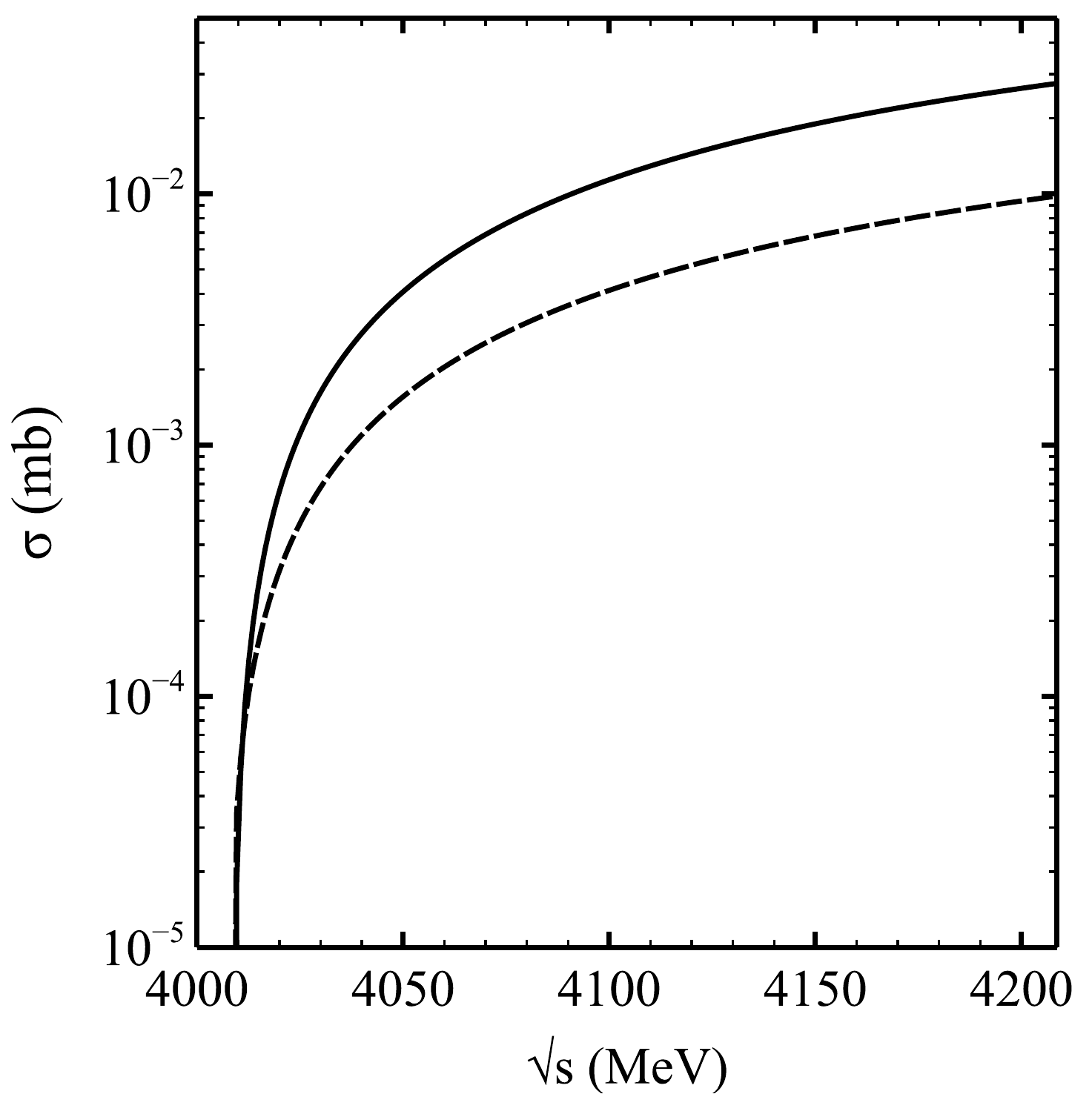}
\caption{Cross section for the reaction $\bar D D\to\pi X$ considering only the neutral components of X (dashed-line) and adding the charged components (solid line).}\label{crossDbarD}
\end{figure}

 In Fig.~\ref{crossDbarD} we show the results obtained for the production cross section of $X$ from the reaction $\bar D D\to\pi X$ as a function of the center of mass energy, $\sqrt{s}$. The dashed line corresponds to the case where only the neutral
components of $X$, i.e, $\bar D^0 D^{*0}-\textrm{c.c}$, are considered in the calculations, as in Ref.~\cite{Cho3}. The solid line is the result for the cross section when all components of $X$ are taken into account (using the couplings shown in Table~\ref{tableX}).
As can be seen from Fig.~\ref{crossDbarD}, the difference between the two curves is around a factor 2-3, depending on the energy. Thus, in a model in which $X$ is considered as a molecular state of $\bar D D^*-\textrm{c.c}$, a precise determination of the magnitude of the production cross section for $X$ necessarily implies the consideration of all the components, neutral as well as charged.

\begin{figure}
\centering
\includegraphics[width=0.4\textwidth]{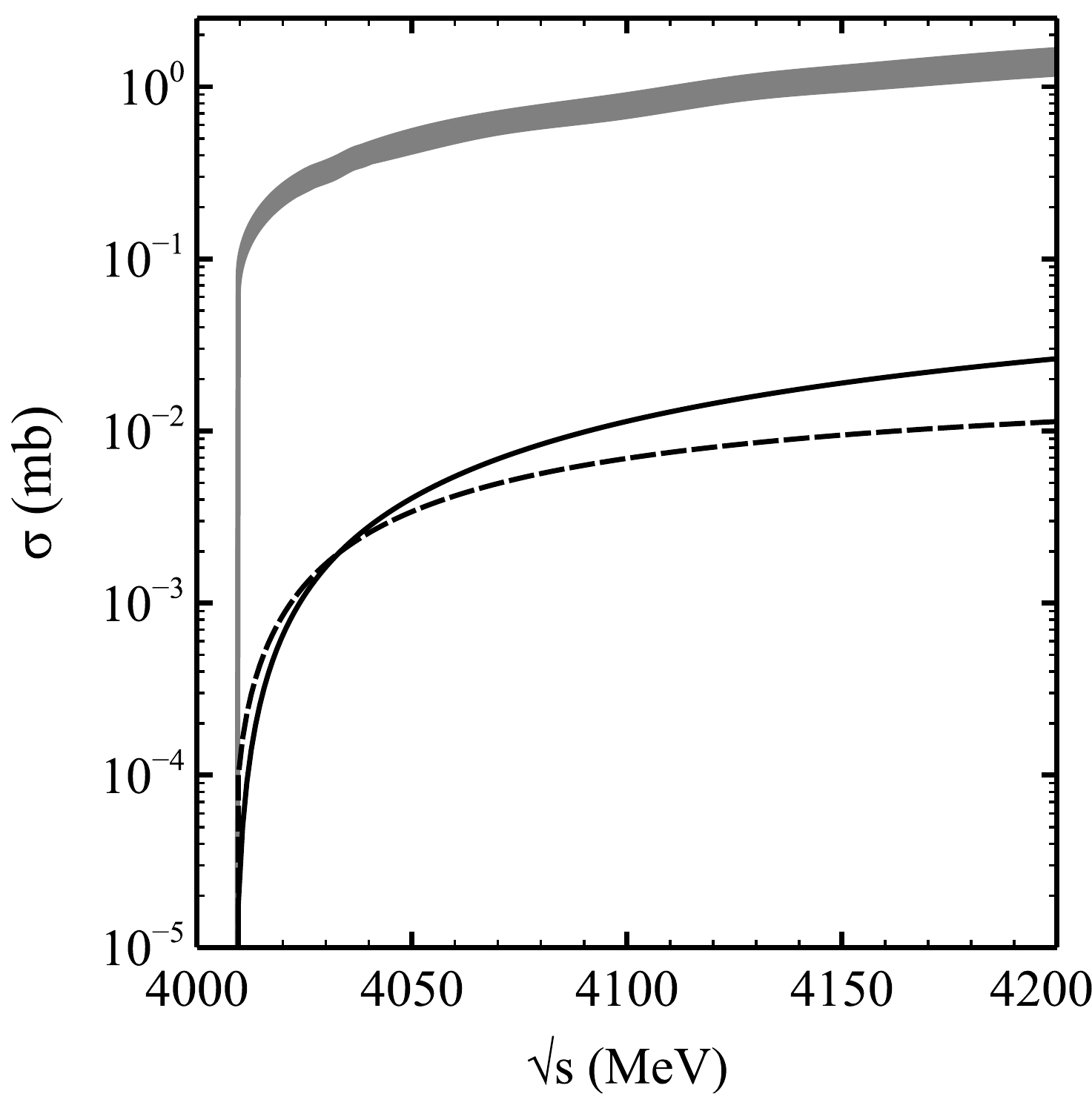}
\caption{Cross section for the reaction $\bar D^* D\to\pi X$. The solid line has the same meaning as in Fig.~\ref{crossDbarD}, and we have shown it for the purpose of comparison. The dashed line represents the result for the cross section of the process $\bar D^* D\to\pi X$ considering only the $t$ channel diagram in Fig.~\ref{DbarstarD}. The shaded region is the result obtained with both $t$ and $u$ channel diagrams of Fig.~\ref{DbarstarD} considering cut-offs in the range 700-1000 MeV.}\label{crossDbarstarD}
\end{figure}

\subsection{The $\bar D^* D\to \pi X$ reaction considering triangular loops}
Next, we determine the cross section related to the process $\bar D^* D\to \pi X$. The diagrams considered for this process (see Figs.~\ref{DbarstarD}c and~\ref{DbarstarD}d) involve anomalous vertices, $\bar D^* \bar D^*\pi$ in the $t$-channel and $X\bar D^* D^*$ in the $u$-channel. We find it interesting to compare the contributions arising form these vertices. We show the results in
Fig.~\ref{crossDbarstarD}. The solid line, as in Fig.~\ref{crossDbarD}, continues representing the final result for the $\bar D D\to \pi X$ cross section. The dashed line is the cross section for the $\bar D^* D\to \pi X$ process without considering the diagrams involving the anomalous vertex $X\bar D^* D^*$, i.e., only with the $t$ channel diagram shown in Fig.~\ref{DbarstarD}c. The shaded region represents the result found with both $t$ and $u$ channel diagrams shown in Figs.~\ref{DbarstarD}c and ~\ref{blob} (with the latter ones involving the $X\bar D^* D^*$ vertex) when changing the cut-off needed to regularize the loop integrals in the range 700-1000 MeV. As can be seen, the results do not get very affected by a reasonable change in the cut-off. Clearly, the vertex $X\bar D^* D^*$ plays an important role in the determination of the $\bar D^* D\to \pi X$ cross section, raising it by around a factor 100-150.

The importance of the anomalous vertices has been earlier mentioned in different contexts. For example, in Ref.~\cite{Oh} the $J/\psi$ absorption cross sections by $\pi$ and $\rho$ mesons were evaluated for several processes producing $D$ and $D^*$ mesons in the final state. The authors found
that  the $J/\psi\,\pi\to  D^*\bar D$ cross section obtained with the exchange of a $D^*$ meson in the $t$-channel, which involves the anomalous $D^* D^*\pi$ coupling, was around 80 times bigger than the one obtained with a $D$ meson exchange in the $t$-channel. In Ref.~\cite{Nagahiro1} the authors studied the radiative decay modes of the $f_0(980)$ and $a_0(980)$ resonances, finding
that the diagrams involving anomalous couplings were quite important for most of the decays, particularly for the $f_0(980)\to \rho^0\gamma$, $a_0(980)\to \rho\gamma$ and $a_0(980)\to\omega\gamma$.

Summarizing this subsection, we have shown that the cross section for the reaction $\bar D^* D\to\pi X$ is larger than that for $\bar D D\to\pi X$ and, thus, the consideration of this reaction in a calculation of the abundance of the $X$ meson in heavy ion collisions could be important.

\subsection{Estimating the $g_{X\bar D^* D^*}$ coupling}

\begin{figure}
\centering
\includegraphics[width=0.4\textwidth]{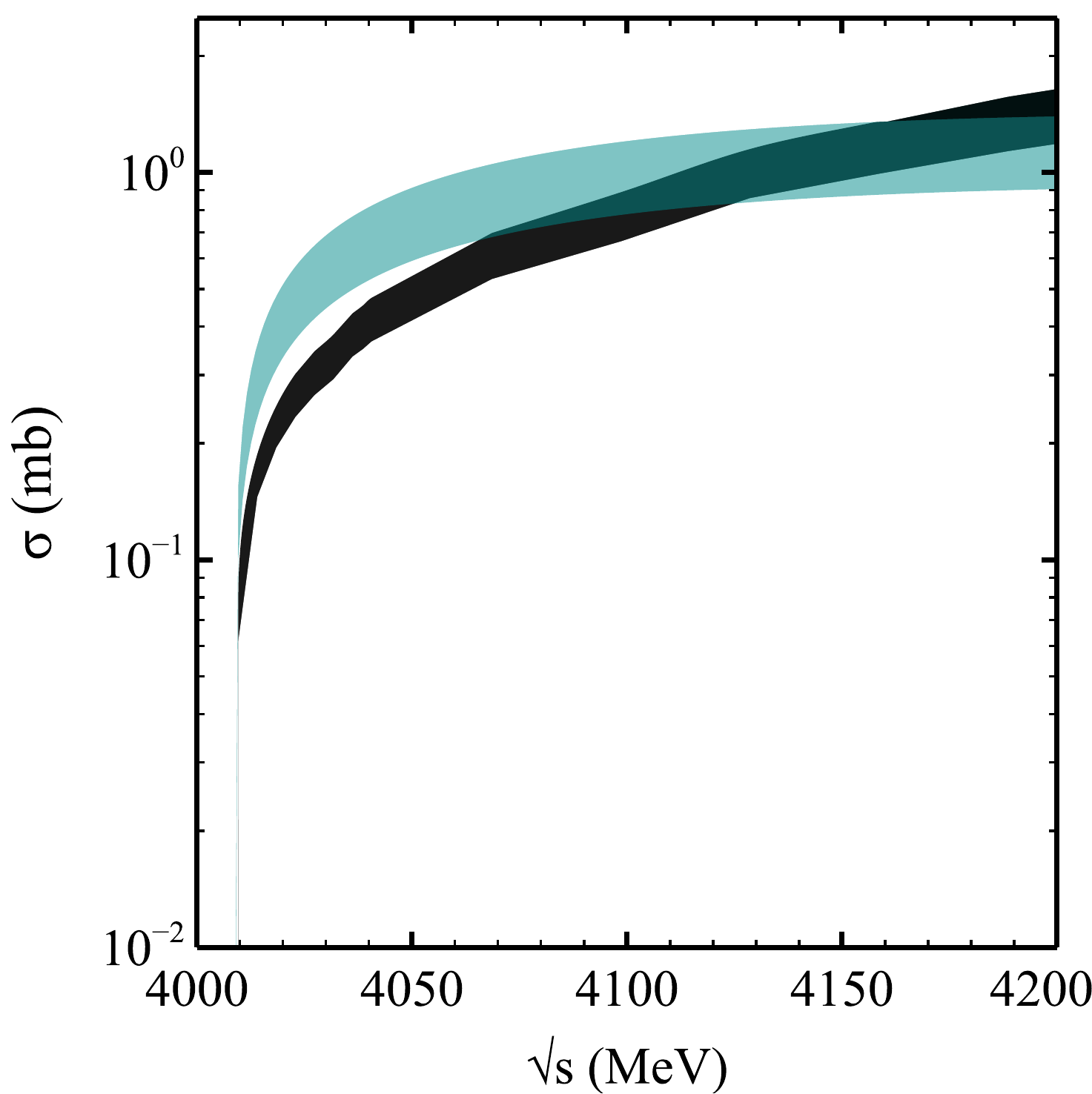}
\caption{Cross section for the reaction $\bar D^* D\to\pi X$. The dark color shaded region has the same meaning as the shaded region in Fig.~\ref{crossDbarstarD}. The light color shaded region represents the result for the cross section when considering the Lagrangian in Eq.~(\ref{effL}) to determine the $X\bar D^* D^*$ vertex with the value of the coupling given in Eq.~(\ref{coupa}).}\label{comp}
\end{figure}

Having determined the contribution from the anomalous vertex $X\bar D^* D^*$ calculating the loops shown in Fig.~\ref{blob}, we could now obtain the cross section for the $\bar D^* D\to\pi X$ reaction using the Lagrangian of Eq.~(\ref{effL}) to determine the amplitude for the diagram shown in Fig.~\ref{DbarstarD}d, which results in Eq.~(\ref{TUDstar}). In this way we can fix the $X\bar D^* D^*$ coupling to that value which gives similar results to the shaded region shown in Fig.~\ref{crossDbarstarD}. From Eq.~(\ref{effL}), it can be seen that the coupling $g_{X\bar D^* D^*}$ should be dimensionless. In Fig.~\ref{comp} we show the results found for the cross section of the reaction $\bar D^* D\to\pi X$ for  $g_{X\bar D^* D^*}$ in the range $1.95\pm 0.22$ (light color shaded region). The dark shaded region in the figure corresponds to the result for the cross section obtained by evaluating the vertex $X\bar D^* D^*$ using the diagrams in Fig.~\ref{blob}, where the loops have been regularized with a cut-off in the range $700-1000$ MeV. It can be seen that, although the energy dependence obtained by using the Lagrangian in Eq.~(\ref{effL}) is not exactly the same as the one found by  considering the triangular loops of Fig.~\ref{blob}, the two results are compatible in some energy range. Thus, the usage of the Lagrangian of Eq.~(\ref{effL}) with the value 
\begin{equation}
g_{X\bar D^* D^*}\sim 1.95\pm 0.22,\label{coupa}
\end{equation}
can be considered as a reasonable approximation for describing processes involving the anomalous vertex $X\bar D^* D^*$, simplifying in this way the calculation of this vertex to a great extend.
\subsection{The $\bar D^* D^*\to\pi X$ reaction}
After estimating the coupling $g_{X\bar D^* D^*}$, we can use this value to determine the cross section for the process 
 $\bar D^* D^*\to\pi X$, which could also get a contribution from the anomalous $X\bar D^* D^*$ vertex, 
that was neglected in Ref.~\cite{Cho3}. The different Feynman diagrams considered for this process are depicted in Fig.~\ref{DstarDstar}.

\begin{figure}[h!]
\centering
\includegraphics[width=0.4\textwidth]{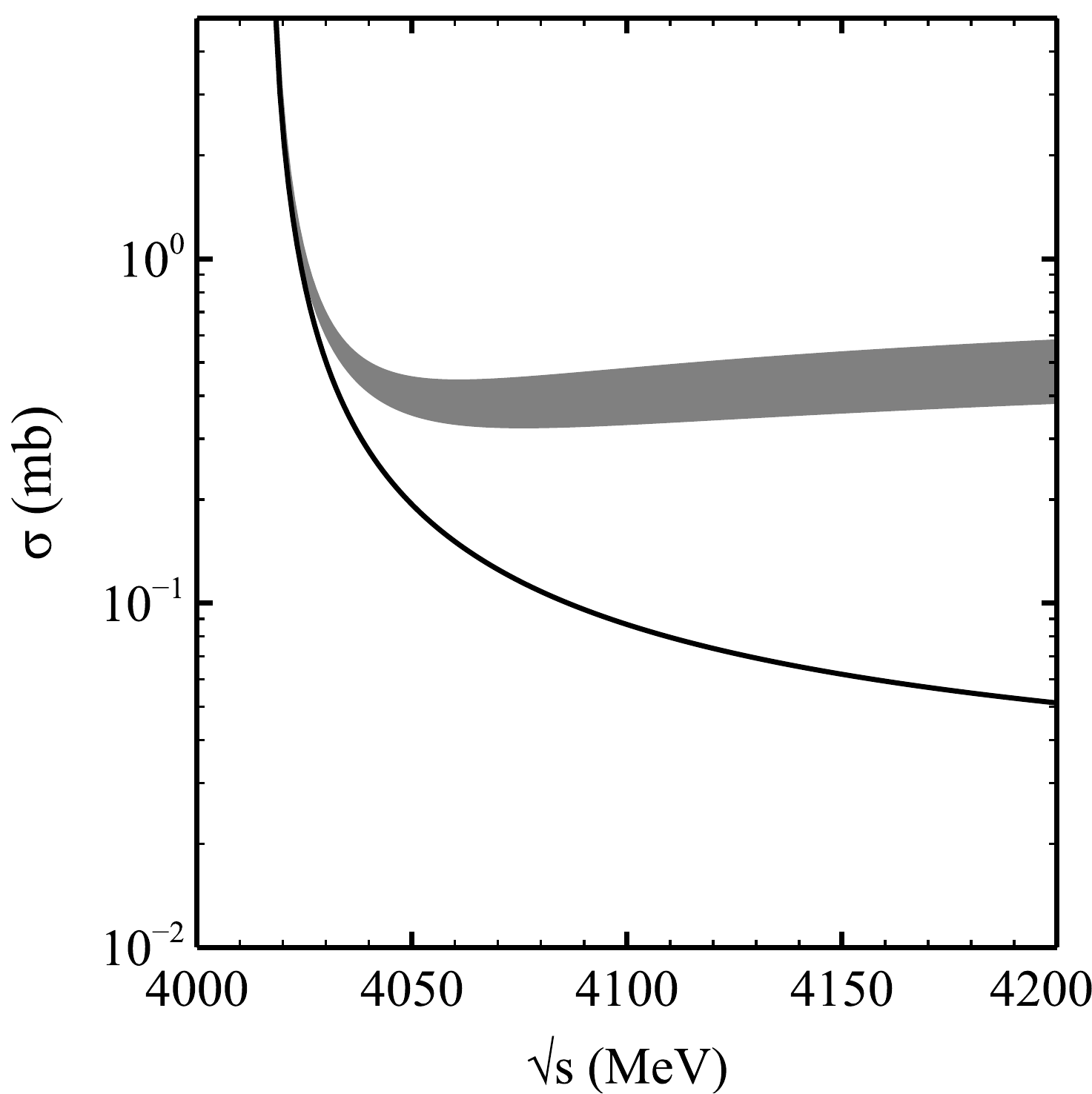}
\caption{Cross section for the reaction $\bar D^* D^*\to\pi X$. The solid line represents the cross section without the contribution from the diagrams in Fig.~\ref{DstarDstar}b and \ref{DstarDstar}d, which contain the vertex $X\bar D^* D^*$. The shaded region represents the result for the cross section when including the contribution of all the diagrams in Fig.~\ref{DstarDstar}, with the vertex $X\bar D^*D^*$ obtained using the Lagrangian in Eq.~(\ref{effL}) with the value of the coupling given in Eq.~(\ref{coupa}).}\label{figDbarstar}
\end{figure}

In Fig.~\ref{figDbarstar} we show the results for the cross section of the reaction $\bar D^* D^*\to \pi X$. The solid line corresponds to the result found without the anomalous $X\bar D^* D^*$ contribution, while the shaded region is the result considering the diagrams involving this anomalous vertex with the value for the $g_{X\bar D^* D^*}$ coupling given in Eq.~(\ref{coupa}).
The first observation to be made is that the cross section for $\bar D^* D^*\to \pi X$ diverges close to the threshold of the reaction. This behavior is different to the cross sections of the processes studied in the previous sections. This is because the reaction $\bar D^* D^*\to \pi X$ is exothermic, while $\bar D D, \bar D^* D\to \pi X$ are endothermic. The second observation is that the contribution from the diagrams involving the $X\bar D^* D^*$ vertex is important, raising the cross section about a factor 8-10. 

Therefore, as in case of the $\bar D^* D\to\pi X$ reaction, the consideration of the anomalous vertices could play an important role when determining the $X$ abundance in heavy ion collisions.

\section{Summary}\label{Sum}

In this work we have obtained the production cross sections of the reactions $\bar D D \to\pi X$, $\bar D^* D \to\pi X$ and   $\bar D^* D^*\to\pi X$, considering $X(3872)$ as a molecular state of $\bar D D^*-\textrm{c.c}$. 
We have shown that the consideration of the neutral as well as the charged hadrons coupling to $X$ is important for the 
evaluation of the cross sections. Next, to obtain the cross section for the process $\bar D^* D\to\pi X$ we have included the 
contribution of the anomalous vertex $X\bar D^* D^*$. With this result, we have estimated the $X\bar D^* D^*$ coupling and used 
it to calculate the cross section for the reaction $\bar D^* D^*\to\pi X$. The contribution to the cross section from the 
vertex $X\bar D^* D^*$ turns out to be important and could play an important role in the determination of the abundance of 
the $X$ meson in heavy ion collisions.

Our results pave the way for a new round of calculations of $X$ abundancies in a 
hadron gas, as outlined in Ref. \cite{Cho3}. We emphasize that 
we expect to find some significant differences with respect to the results found in Ref.~\cite{Cho3}, because the processes 
 $\bar D D \to\pi X$ and  $\bar D^* D^* \to\pi X$ have been recalculated and, more importantly, the process 
$\bar D^* D \to\pi X$ has been included. This latter was found to give the most important contribution of all the three  
processes considered.

\section{Acknowledgements}
The authors would like to thank the Brazilian funding agencies FAPESP and CNPq for the financial support. \\


\begin{thebibliography}{99}
\bibitem{Uchida} 
  M.~Uchida [Belle Collaboration],
  Few Body Syst.\  {\bf 54}, 947 (2013).
\bibitem{Guo} 
  Y.~p.~Guo [BESIII Collaboration],
  EPJ Web Conf.\  {\bf 72}, 00009 (2014).
  \bibitem{Choi} 
  S.~K.~Choi {\it et al.}  [Belle Collaboration],
  Phys.\ Rev.\ Lett.\  {\bf 91}, 262001 (2003).
 
 \bibitem{Acosta} 
  D.~Acosta {\it et al.}  [CDF Collaboration],
  Phys.\ Rev.\ Lett.\  {\bf 93}, 072001 (2004).
  
  \bibitem{Abazov} 
  V.~M.~Abazov {\it et al.}  [D0 Collaboration],
  Phys.\ Rev.\ Lett.\  {\bf 93}, 162002 (2004).
  
  \bibitem{Aubert} 
  B.~Aubert {\it et al.}  [BaBar Collaboration],
  Phys.\ Rev.\ D {\bf 71}, 071103 (2005).

   \bibitem{Aaij} 
  R. Aaij {\it et al.}  [LHCb Collaboration],
  Phys.\ Rev.\ Lett.\  {\bf 110}, no. 22, 222001 (2013).
  
    \bibitem{Tornqvist} 
  N.~A.~Tornqvist,
  Phys.\ Lett.\ B {\bf 590}, 209 (2004).
  
\bibitem{Close} 
  F.~E.~Close and P.~R.~Page,
  Phys.\ Lett.\ B {\bf 578}, 119 (2004).
    
  \bibitem{Swanson} 
  E.~S.~Swanson,
  Phys.\ Lett.\ B {\bf 588}, 189 (2004).
  
  \bibitem{Braaten} 
  E.~Braaten and M.~Kusunoki,
  Phys.\ Rev.\ D {\bf 72}, 054022 (2005).
  
  \bibitem{Daniel2}
  D. Gamermann and E. Oset, Eur. Phys. J. A {\bf 33}, 119 (2007).
  
\bibitem{Nielsen} 
 R.~D'E.~Matheus, S.~Narison, M.~Nielsen and J.~M.~Richard,
  Phys.\ Rev.\ D {\bf 75}, 014005 (2007).
 
 \bibitem{Matheus} 
  R.~D'E.~Matheus, F.~S.~Navarra, M.~Nielsen and C.~M.~Zanetti,
  Phys.\ Rev.\ D {\bf 80}, 056002 (2009).
  
  \bibitem{Dong} 
  Y.~Dong, A.~Faessler, T.~Gutsche, S.~Kovalenko and V.~E.~Lyubovitskij,
  Phys.\ Rev.\ D {\bf 79}, 094013 (2009).
  
    \bibitem{Daniel3}
   D. Gamermann and E. Oset, Phys. Rev. D {\bf 80}, 014003 (2009).
  
    
  \bibitem{Daniel4} 
  D.~Gamermann, J.~Nieves, E.~Oset and E.~Ruiz Arriola,
  Phys.\ Rev.\ D {\bf 81}, 014029 (2010).
  
   \bibitem{Dubnicka} 
  S.~Dubnicka, A.~Z.~Dubnickova, M.~A.~Ivanov, J.~G.~Koerner, P.~Santorelli and G.~G.~Saidullaeva,
  Phys.\ Rev.\ D {\bf 84}, 014006 (2011).
  
  \bibitem{Badalian} 
  A.~M.~Badalian, V.~D.~Orlovsky, Y.~.A.~Simonov and B.~L.~G.~Bakker,
  Phys.\ Rev.\ D {\bf 85}, 114002 (2012).
      
   \bibitem{Coito} 
  S.~Coito, G.~Rupp and E.~van Beveren,
  Eur.\ Phys.\ J.\ C {\bf 71}, 1762 (2011), {\it idem} Eur.\ Phys.\ J.\ C {\bf 73}, 2351 (2013).
  
  \bibitem{roma14}  
  A.~L.~Guerrieri, F.~Piccinini, A.~Pilloni and A.~D.~Polosa,
  Phys.\ Rev.\ D {\bf 90}, 034003 (2014).


\bibitem{Arsene}   I.~Arsene {\it et al.}  [BRAHMS Collaboration],
                   Nucl.\ Phys.\ A {\bf 757}, 1 (2005).
  
\bibitem{Adams}    J.~Adams {\it et al.}  [STAR Collaboration],
                   Nucl.\ Phys.\ A {\bf 757}, 102 (2005).
                   
 \bibitem{Cho1} 
  S.~Cho {\it et al.}  [ExHIC Collaboration],
  Phys.\ Rev.\ Lett.\  {\bf 106}, 212001 (2011).
  
  \bibitem{Cho2} 
  S.~Cho {\it et al.}  [ExHIC Collaboration],
  Phys.\ Rev.\ C {\bf 84}, 064910 (2011).

   \bibitem{Cho3} 
  S.~Cho and S.~H.~Lee,
  Phys.\ Rev.\ C {\bf 88}, 054901 (2013).
  
  \bibitem{Oh}   Y.~S.~Oh, T.~Song and S.~H.~Lee,
  Phys.\ Rev.\ C {\bf 63}, 034901 (2001).
 \bibitem{Nagahiro1} 
  H.~Nagahiro, L.~Roca and E.~Oset,
  Eur.\ Phys.\ J.\ A {\bf 36}, 73 (2008).
  
  \bibitem{Nagahiro2} 
  H.~Nagahiro, L.~Roca, A.~Hosaka and E.~Oset,
  Phys.\ Rev.\ D {\bf 79}, 014015 (2009).
  
  \bibitem{Ozaki} 
  S.~Ozaki, H.~Nagahiro and A.~Hosaka,
  Phys.\ Lett.\ B {\bf 665}, 178 (2008).
  
  \bibitem{Francesca2} 
  F.~Aceti, R.~Molina and E.~Oset,
  Phys.\ Rev.\ D {\bf 86}, 113007 (2012).
   \bibitem{Liang}
  W.~H.~Liang, C.~W.~Xiao and E.~Oset,
  Phys.\ Rev.\ D {\bf 89}, 054023 (2014).
  
  \bibitem{Francesca} 
  F.~Aceti, M.~Bayar and E.~Oset,
  Eur.\ Phys.\ J.\ A {\bf 50}, 103 (2014).

  
  
  \bibitem{Anastassov} 
  A.~Anastassov {\it et al.}  [CLEO Collaboration],
  Phys.\ Rev.\ D {\bf 65}, 032003 (2002).
  
  \bibitem{Bracco} 
  M.~E.~Bracco, M.~Chiapparini, F.~S.~Navarra and M.~Nielsen,
  Prog.\ Part.\ Nucl.\ Phys.\  {\bf 67}, 1019 (2012).

\bibitem{our} 
  A.~Martinez Torres, K.~P.~Khemchandani, F.~S.~Navarra, M.~Nielsen and L.~M.~Abreu,
  arXiv:1405.7583 [hep-ph].
    \bibitem{Bando1} 
  M.~Bando, T.~Kugo, S.~Uehara, K.~Yamawaki and T.~Yanagida,
  Phys.\ Rev.\ Lett.\  {\bf 54}, 1215 (1985).
  
  \bibitem{Bando2} 
  M.~Bando, T.~Kugo and K.~Yamawaki,
  Phys.\ Rept.\  {\bf 164}, 217 (1988).
  
  \bibitem{Meissner} 
  U.~G.~Meissner,
  Phys.\ Rept.\  {\bf 161}, 213 (1988).
  
  \bibitem{Harada} 
  M.~Harada and K.~Yamawaki,
  Phys.\ Rept.\  {\bf 381}, 1 (2003).
  
  \bibitem{Maiani}  
  L.~Maiani, F.~Piccinini, A.~D.~Polosa and V.~Riquer, 
  Phys.\ Rev.\ D {\bf 71}, 014028 (2005). 
     
\end{thebibliography}
\end{document}